\documentstyle[prb,aps]{revtex}

\begin{document}
\title{A study of the mechanisms of the semi-insulating conversion of InP by
anelastic spectroscopy}
\author{R. Cantelli}
\address{Universit\`{a} di Roma ``La Sapienza``, Dipartimento di Fisica, P.le A.\\
Moro 2, I-00185 Roma, and INFM, Italy}
\author{F. Cordero}
\address{CNR, Area di Ricerca di Tor Vergata, Istituto di Acustica ``O.M. Corbino``,\\
Via del Fosso del Cavaliere 100, I-00133 Roma, and INFM, Italy}
\author{O. Palumbo}
\address{Universit\`{a} di Roma ``La Sapienza``, Dipartimento di Fisica, P.le A.\\
Moro 2, I-00185 Roma, and INFM, Italy}
\author{G. Cannelli}
\address{Universit\`{a} della Calabria, Dipartimento di Fisica,\\
Arcavacata di Rende (CS),I-87036 Cosenza, and INFM, Italy}
\author{F. Trequattrini}
\address{Universit\`{a} di Roma ``La Sapienza``, Dipartimento di Fisica, P.le A.\\
Moro 2, I-00185 Roma, and INFM, Italy}
\author{G.M. Guadalupi, B. Molinas}
\address{Venezia Tecnologie SpA (ENI Group), Via delle Industrie 39,\\
I-30175 P. Marghera (VE), Italy}
\maketitle

\begin{abstract}
Elastic energy absorption measurements versus temperature on semiconducting,
semi-insulating (SI) and Fe-doped InP are reported. A thermally activated
relaxation process is found only in the SI state, which is identified with
the hopping of H atoms trapped at In vacancies. It is proposed that the
presence of In vacancies in InP prepared by the liquid encapsulated
Czochralski method is due to the lowering of their energy by the saturation
of the P dangling bonds with H atoms dissolved from the capping liquid
containing H$_{2}$O. The conversion of iron-free InP to the SI state
following high temperature treatments would be due to H loss with the
transformation of the H-saturated In vacancies, V$_{\text{In}}$-H$_{4}$
donors, into neutral and acceptor V$_{\text{In}}$-H$_{n}$ complexes with $%
n<4 $. Such complexes would produce the observed anelastic relaxation
process and may also act as deep acceptors which neutralize unwanted donor
impurities.
\end{abstract}

\twocolumn

\section{INTRODUCTION}

Semi-insulating (SI) InP is usually obtained by adding Fe before growing the
crystal by the liquid encapsulated Czochralski (LEC) method. SI InP:Fe
offers remarkable perspectives as a substrate in optical fiber communication
systems, and in high frequency electronic devices. Essential prerequisites
are both high electrical resistivity and carrier mobility, in addition to a
high material homogeneity. Generally, the semiconducting as-grown InP is
quite inhomogeneous, as it exhibits growth striations, dislocations
decorated by microdefects, microprecipitates embedded in the crystal lattice
and sometimes inclusions.\cite{AJA97} In particular, since the Fe
concentrations necessary to reach the semi-insulating state are close the Fe
solubility limit, domains of Fe precipitates are inevitably formed.
Post-growth thermal treatments of the ingots at about 900$~^{\text{o}}$C
produce beneficial effects concerning defect annealing besides a resistivity
increase, but cause inhomogeneities across the crystal diameter\cite{ZFC97}
and do not improve the Fe homogeneity.\cite{SGA97} Therefore, the thermal
treatments have to be carried out on the wafers cut from the as-grown
ingots; in this way the wafer homogeneity results improved\cite{FGS97} and
the striations and other types of lattice imperfections may be annealed.

It is diffusely believed that in the SI state of InP:Fe, the electrons
supplied to the crystal by the unwanted residual impurities (shallow donors)
are trapped by Fe which acts as a deep acceptor and provides compensation.
However, the SI state is reached only if the Fe concentration (typically in
the range of 10$^{16}$~cm$^{-3}$) is remarkably higher than the residual
donor impurity content; as a confirmation, it was reported that the impurity
content in SI InP:Fe is lower than the free carrier concentration,\cite
{Zac93} indicating that sources of donors other than the impurities
considered above must be active; they were identified later as V$_{\text{In}%
} $(HP)$_{4}$ complexes,\cite{BBG95,DPR93} and labelled here as V$_{\text{In}%
}- $H$_{4}$. Lastly, it should be mentioned that the achievement of the SI
state by Fe doping may present further disadvantages due to the
nonuniformity of the electrical properties along the crystal growth axis
caused by Fe segregation.\cite{SMF93}

The semi-insulating state is also reached by thermally treating undoped InP
for several hours or days at about 950$~^{\text{o}}$C under phosphorous
atmospheres ranging from tenths to tens of bars,\cite{KHK96,HMS89,KSY91}
after which the resistivity can be increased above 10$^{5}$~$\Omega $~cm.
However, for the conversion to occur it is necessary that the carrier
concentrations be below the limit of 4x10$^{15}$~cm$^{-3}$ (Ref. %
\onlinecite{HMS89}); thermal treatments of materials having higher residual
carrier concentrations due to higher impurity contents in the starting
material usually display only a slight reduction in the concentration of
free carriers after annealing,\cite{HWM93} and the same effect is also
produced in SI InP:Fe.\cite{ZFC97}

Several mechanisms have been proposed for the SI conversion, and many of
them hypothesize that the thermal treatment may activate some kinds of deep
acceptors in such a concentration to compensate the shallow donors related
to impurities. Among the possible deep acceptors, the Fe atom both in
Fe-doped and in undoped InP (therefore present as residual impurity) has
been considered. However, also the hydrogen-related defects are receiving
great attention, as H is always present\cite{CCN87} and may form complexes
with native defects and impurities.

Generally, the studies of the properties of SI InP are conducted on the
material doped with Fe, whilst a systematic analysis of the comparative
properties of SI InP:Fe and undoped InP is missing. The purpose of the
present work is to carry out, through measurements of the complex modulus
(elastic energy loss and dynamic modulus), a comparative study of the
defects in InP in the semiconducting and the semi-insulating state obtained
both by the doping with Fe and by the thermal treatment.

\section{EXPERIMENT}

The thermal treatments (TT) finalized to the conversion from the
semiconducting to the SI state were carried out on wafers at about 950~$^{%
\text{o}}$C under a phosphorous vapor pressure of $\sim 0.2$~bar for times
of the order of hours.

The crystals were grown by the LEC method at Venezia Tecnologie, and the
samples were three rectangular bars (about $43\times 5\times 0.4$~mm$^{3}$)
of normal purity{\em \ }and high purity. They were characterized as follows:
sample 1 was an undoped highly pure sample subjected to the TT for
conversion and presented the SI state. Sample 2{\em \ }was of normal purity
and doped with 10$^{16}$~cm$^{-3}$ Fe impurities; therefore it was therefore
semi-insulating without need of TT. Sample 3 was highly pure but not
subjected to the thermal treatment for the conversion, and hence non
semi-insulating.

The elastic energy loss and resonant frequency measurements were carried out
between 1.9 and 650$~$K. The samples were horizontally suspended on two
nodal lines by thin wires and three flexural vibration modes were
electrostatically excited in the frequency range between 1 and 16~kHz. The
vibration amplitude was detected through the variation of the
sample-electrode capacitance by a frequency modulation technique, the
elastic energy absorption was measured from the rate of the free decay of
the vibration.

The elastic energy loss coefficient, or reciprocal of the mechanical $Q$ of
the sample, is related to the imaginary part of the complex Young's modulus $%
E^{*}$ as\cite{NB} 
\begin{equation}
Q^{-1}={\rm Im}\left( \frac{\delta E}{E}\right) =E\,\frac{v_{0}\,c}{k_{\text{%
B}}T}{\rm \,}\left( \Delta \lambda \right) ^{2}\frac{\omega \tau }{1+\left(
\omega \tau \right) ^{2}}\,,
\end{equation}
where the last expression is the contribution of a molar concentration $c$
of defects diffusing or reorienting with a characteristic time $\tau $
between states or sites which are made inequivalent by the vibration stress $%
\sigma $; $\Delta \lambda $ is the appropriate component of the change of
the elastic dipole tensor after a defect reorientation, and the elastic
dipole tensor $\lambda $ is defined as the strain caused by a unit
concentration of the defects; $v_{0}$\ is the cell volume and $\omega =2\pi
f $ is the angular vibration frequency. The above contribution to the
absorption is peaked at the temperature at which $\omega \tau =1$; since $%
\tau $\ is temperature dependent and usually follows an Arrhenius law, $\tau
\left( T\right) =\tau _{0}\exp \left( E/k_{\text{B}}T\right) $, the peak
shifts to higher temperature if measured at higher $\omega $. The present
samples had the $\left\langle 110\right\rangle $ axis parallel to the
length, so that the appropriate modulus is\cite{NB} $E_{\left[ 110\right]
}^{-1}=\frac{1}{2}\left( s_{11}+s_{12}+\frac{1}{2}s_{44}\right) $. For
trigonal defects $\delta E_{\left[ 110\right] }^{-1}=\frac{1}{2}\delta
s_{44}=\frac{cv_{0}}{kT}\frac{2}{9}\left( \lambda _{1}-\lambda _{2}\right)
^{2}$ while for $\left\langle 110\right\rangle $ orthorhombic defects $%
\delta E_{\left[ 110\right] }^{-1}=\frac{1}{6}\delta \left(
s_{11}-s_{12}\right) +\frac{1}{2}\delta s_{44}=\frac{cv_{0}}{kT}\frac{1}{6}%
\left[ \frac{1}{12}\left( \lambda _{1}+\lambda _{2}-2\lambda _{3}\right)
^{2}+\left( \lambda _{1}-\lambda _{2}\right) ^{2}\right] $. The Young's
modulus is related to the frequency of the first flexural mode by $E=\rho
\,(0.975\,f\,\,l^{2}/t)^{2}$, where $\rho $, $l$ and $t$ are the sample
density, length and thickness; for our samples it is $E_{\left[ 110\right]
}=1.40\times 10^{11}$~Pa.

\section{RESULTS}

Figure 1 shows the elastic energy dissipation curves from 1.9 to 370$~$K of
the undoped SI sample 1 and sample 2. An intense peak (peak A) is present at
about 290$~$K in both samples, whereas the dissipation curves in the low
temperature side indicate that two additional peaks (around 130 and 220$~$K)
may be present but confused with the background; however, due to their small
height, they will not be considered here. It is surprising that the two
samples, having in common only the SI state but not the purity and the
thermal history, present the same type of spectrum, and differ only for the
height of the background dissipation.

In order to test the thermal stability of the anelastic relaxation spectrum,
sample 2 was annealed in vacuum for about 30 min at temperatures
subsequently increased from 400 to 660$~$K at steps of 40$~$K; after each
step the relaxation spectrum was measured, and the results are reported in
Fig. 2. With increasing annealing temperature the background dissipation
progressively decreases down to a saturation value which is attained after
the treatment at 620$~$K (7$^{\text{th}}$). Instead, the height of the peak
remains practically unaffected, and is also stable versus thermal cycling,
as the curves are retraced on heating and cooling. After this series on
annealings, it was checked that the sample remained in the semi-insulating
state.

The energy-loss curves, measured at three vibration modes (between 1 and
16~kHz) showed that the peak A is thermally activated, since it shifts
towards higher temperatures with increasing frequency. Figure 3 presents the
experimental data of sample 2 after the 7$^{\text{th}}$ vacuum annealing at
620$~$K at 1.2 and 15.6~kHz, after subtraction of the background; the data
are satisfactorily fitted by Debye curves (continuous lines) with a single
relaxation time following the classical Arrhenius law. The activation energy
of the relaxation process and the pre-exponential factor of the relaxation
rate obtained from the fit are $E=0.67~$eV and $\tau _{0}^{-1}=3\times
10^{14}$~s$^{-1}$.

The spectrum of the semiconducting sample 3 is reported in Fig. 4 together
with those of samples 1 and 2 (continuous lines) for comparison. Sample 3
displays (curve 1), superimposed to a high background, a small peak at 220$~$%
K and a well developed peak at 400$~$K which decreases during its
measurement and disappears after a vacuum annealing of about 30 min at 450$~$%
K (curve 2). It is remarkable that the peak A, found in the SI samples, is
absent here, or at least non emerging from the background. The lack of peak
A in the non-SI samples was confirmed also in undoped InP samples cut from
the same ingot and from a different ingot (results not shown).

In order to study the stability of the anelastic spectrum in the non-SI
sample 3, a series of vacuum annealings of 30 min each at increasing
temperatures was conducted, as for sample 2. The results are displayed in
Fig. 5 and show that: {\it i}) the dissipation around 400$~$K is not a
stable peak as a function of temperature; indeed, it starts decreasing on
heating above 350$~$K, even keeping the sample temperature constant (result
not reported here) and practically disappears after a vacuum annealing at 500%
$~$K; {\it ii}) the small peak around 220$~$K increases after the first
vacuum annealing (450$~$K), but is depressed after the subsequent treatments
at higher temperatures; {\it iii}) the overall dissipation steadily
decreases down to saturation values with the subsequent isochronal
annealings at progressively higher temperatures.

Figure 6 compares the dissipation curves of SI InP:Fe (sample 2) and of
non-SI InP (sample 3) after the last annealing treatments at 650$~$K; only
the peak at 300$~$K in the SI sample persists after the high temperature
treatments.

In order to test whether peak A is suppressed by heating above 660$~$K, the
isochronal annealings on sample 2 were continued up to 847$~$K, and Fig. 7
shows that the species giving rise to peak A is stable up to that
temperature. At this temperature a few droplets appeared on the sample
surface, indicating surface degradation due to P loss and consequent
formation of In clusters.\cite{Pea94} However, the sample bulk was very
little affected, as deduced by the small relative variation of the resonance
frequency ($<3.5\times 10^{-3}$); the vibration frequency is directly
related to the elastic modulus, which is very sensitive to the formation of
second phases. In conclusion, peak A is stable up to the sample
decomposition in vacuum.

To better interpret the annealing behavior observed, two samples cut from
the same wafers of sample 2 (Fe-doped) and sample 3 (non-SI) were subjected
to vacuum effusion experiments with a rate of 3-4$~$K/min, monitored with a
residual gas analyzer. It was found that H outdiffusion starts above 560$~{^{%
\text{o}}}$C, i.e. at about the same temperature at which the sample
decomposition starts, as indicated by the appearance of some peaks in the
effusion spectrum corresponding to the masses of PH$_{n}$ and P$_{2}$H$_{n}$%
, besides H$_{2}$.

\section{DISCUSSION}

\subsection{The vacancy-H complexes}

Optical absorption experiments on InP:Fe have revealed several local
vibrational modes (LVM), regardless of whether the material was as-grown,%
\cite{BBG95,Paj90} plasma hydrogenated,\cite{DPR93,CCP91} or ion implanted:%
\cite{TCC88,RNS88,FMM92} these lines occur in the range included between
2200 and 2300~cm$^{-1}$ of the spectral region. However, if the same
material is implanted with deuterons\cite{FMM92} instead of protons, the
corresponding LVM are shifted to lower energies ($\sim 1600$~cm$^{-1}$), and
the ratio between the energies is close to $(m_{\text{D}}/m_{\text{H}%
})^{1/2} $. This observation provided evidence that the peaks are due to
hydrogen related defects cause, and they have been attributed to PH
stretching modes.\cite{BBG95,DPR93,Paj90,CCP91,TCC88,RNS88,FMM92}

Four of the reported lines are particularly important, as they have been
reproducibly observed, and there seems to be an agreement on their
attribution; such peaks occur at 2202~cm$^{-1}$ (peak 1), 2273~cm$^{-1}$
(peak 2), 2286~cm$^{-1}$ (peak 3), and 2316~cm$^{-1}$ (peak 4). The main
line is peak 4 (Ref. \onlinecite{DPR93}), but peak 1 is of comparable
intensity if the samples are ion implanted;\cite{FMM92} instead, peaks 2 and
3 are rather weak.

Peak 1 is originated by a H complex with trigonal symmetry with a P-H bond
oriented along the $\left\langle 111\right\rangle $ direction;\cite
{DPR93,CCP91} the low energy of its stretching mode is indicative of a long
bond length: therefore it has been attributed to an In vacancy having one
the P dangling bonds saturated by a H atom occupying a bond centre (BC) site
between V$_{\text{In}}$ and P; this complex will be labelled as V$_{\text{In}%
}$-H.

Peak 4 appears in the high energy region of the PH stretching mode,
indicating a shorter bond length. Optical absorbance measurements in undoped
InP samples cocharged with H and D have convincingly demonstrated that the
complex giving rise to this peak is constituted by more than one H atom.\cite
{BBG95} In addition, the stress splitting of this line indicates a cubic
symmetry for the corresponding cluster.\cite{DPR93} The various authors
agree that the complex consists of an In vacancy decorated by four H atoms
in BC occupancy, V$_{\text{In}}-$H$_{4}$; in this configuration, the
repulsion between the four H atoms shortens the P-H bond lengths and gives
rise the observed high energy LVM. The four H atoms, each bonded to one of
the four P atoms around V$_{\text{In}}$, contribute four electrons to the
trivalent vacancy and leave one electron available.\cite{BBG95} The V$_{%
\text{In}}-$H$_{4}$ complex is therefore a shallow donor, and could be
responsible for the discrepancy between the donor content estimated from the
concentrations of the impurities and from the Fe$^{+}$ content that is
necessary to passivate them.\cite{BBG95}

Peak 3 was studied by local vibrational mode experiments and was attributed
to the Zn residual impurities;\cite{DPR93,CCN87A} the complex is believed to
be formed by the Zn atom occupying an In site having one H atom as a nearest
neighbor bonded to P. Peak 2 is of uncertain origin, but may be due to a
cluster similar to that of peak 3 with Mn.\cite{CCP91}

\subsection{Peak A}

In InP, the omnipresence of hydrogen\cite{CCN87} (certainly due to the LEC
preparation method), the fact that LVM lines due to isolated H have not been
reported, the ability of H to form complexes\cite
{BBG95,DPR93,Paj90,CCP91,TCC88,RNS88,FMM92} and its high mobility, strongly
indicate that the atomic defect giving rise to peak A may be a H-related
complex. Undoubtedly, the presence of the thermally activated process with
maximum near 300$~$K proves that the InP samples in the SI state contain a
species which is highly mobile at room temperature ($\sim 10^{3}$ jumps per
second) and that this species is rather abundant, considering the height of
the corresponding peak. In addition, the value of the pre-exponential factor
is typical of point defect relaxations and indicates that this imperfection
involves an atomic complex.

It is remarkable that the process is characterized by a single relaxation
time: in fact, the absence of peak broadening indicates that the mobile
species relaxes with only one type of elementary jump, and that there is no
interaction among the relaxing units. Single-time Debye processes are not
common, and one was also discovered in Si:B charged with hydrogen.\cite
{CCC91} In that system the H atom hops among the four equivalent BC sites
coordinated with substitutional B, overcoming an energy barrier ($E=0.22$%
~eV) which is lower than that presently reported. The absence of broadening
of the peak due to the reorientation of the H-B pair led to the conclusion
that the interaction between H-related defects may be negligible at dopant
concentrations as high as 10$^{19}$~at/cm$^{3}$. The relaxational dynamics
of H within the H-B pairs in Si has also been measured by the decay of the
stress-induced IR dichroism\cite{SBP88} for relaxation times between 10$^{3}$
and 10$^{6}$~s. By joining the data for the two types of experiments,\cite
{CS94,CCC96,CC98} relaxation rates spanning 11 orders of magnitude were
obtained in the temperature range 60-120$~$K; a weak deviation from the
Arrhenius dependence was observed at low temperature, which was interpolated%
\cite{CS94} with a polaron-like model.

Similarly to the case of B-H in Si, it would be natural to think of the
reorientation of H around Fe as a source of anelastic relaxation in the
Fe-doped InP sample. Indeed, measurements of the local vibrational modes
show that in InP:Fe, H resides between the Fe atom (which substitutes In)
and a nearest neighbor P atom, mainly bonded to P. However, the fact that
both undoped and Fe-doped InP display the same peak (see for instance Fig.
1, where the small temperature shift between the two peaks is due to the
different frequencies) demonstrate that Fe does not play any role in the
mechanism causing peak A.

We discuss now the possibility that peak A is due to V$_{\text{In}}-$H$_{n}$
complexes ($n=0-4$). For $n=0$ and $n=4$ the complexes have in principle the
same cubic symmetry of the crystal, so that there would not be any
differentiation between complexes under stress, unless Jahn-Teller type
lowering of the defect symmetry occurs. Indeed, it has been proposed that in
certain ionization states the V$_{\text{In}}$ is distorted.\cite{SVP94} In
this case, different distortions would be distinguished by stress and would
cause anelastic relaxation, but the reorientation between them would be
expected to occur with a faster rate than in the cases where atomic hopping
is involved and produce a peak at much lower temperature:\cite{NB} the
barrier of $0.67~$eV seems too high for the reorientation of a Jahn-Teller
like distortion.

Instead, the reorientation of the V$_{\text{In}}-$H$_{n}$ complexes with $%
n=1,2$ and $3$ is expected to appear as a slower single-time relaxation
process. In fact, in all three cases the complexes can reorient among
equivalent configurations through a single elementary jump of an H atom
between the four different P dangling bonds around the vacancy. For $n=1$
there are four equivalent BC sites for H, each coordinated with one of the
four P bonds, and the defect has trigonal symmetry. For $n=3$, one of the
four BC sites is unoccupied, so that the complex reorientation is due to the
jump of the H vacancy instead of the H atom, and the symmetry is again
trigonal. For $n=2$ the H atoms can occupy six equivalent pairs of BC sites
along the [110] directions. The corresponding elastic dipole has the
principal axes along two $\left\langle 110\right\rangle $ and one $%
\left\langle 100\right\rangle $ directions and its symmetry is $\left\langle
110\right\rangle $ orthorhombic.

The three complexes may have different energy barriers for reorientation,
producing distinct relaxation peaks. In this case, peak A would be due to
the dominant complex, and the other two complexes would produce peaks at
different temperatures, e.g. the two small peaks at 130 and 220$~$K. It is
also possible that all the complexes have nearly the same energy barrier for
reorientation. In fact, the LVM energies of the various complexes differ by
only 5\% (Refs. \onlinecite{BBG95,DPR93}), and this can be taken as an
indication that the potential felt by H and therefore the barriers between
different minima are little affected by the presence of neighboring H atoms.
In this case all the three complexes would contribute to peak A, and indeed
the present data are compatible with a spread of activation energies of the
order of 5\%. At present it is not possible to decide between the two cases,
and further measurements on crystals with different orientations are needed.

Finally, we note that the concentration of defects producing peak A
estimated from the peak intensity is fully consistent with the concentration
of H and V$_{\text{In}}$-H$_{n}$ complexes which are generally present in
InP crystals grown by LEC,\cite{CCN91} i.e. of the order of $10^{16}$~cm$%
^{-3}$. In fact, the peak intensity in Eq. (1) is 
\begin{equation}
Q_{\max }^{-1}=c\frac{v_{0}E}{2k_{\text{B}}T}\Delta \lambda ^{2}
\end{equation}
and substituting the measured $E=1.40\times 10^{11}$~Pa and assuming that%
\cite{NB} $0.1\leq \Delta \lambda \leq 1$, the volume concentration of
defects turns out of the order of $c/v_{0}\sim 6\times 10^{16\pm 1}$~cm$^{-3}
$.

\subsection{The background dissipation and the peak at 400$~$K}

The high background dissipation of both the SI (Fig.2) and the
semiconducting (Fig.5) samples might be due to the thermoelastic effect or
to the movement of dislocations. The thermoelastic effect is due to the
periodic heat transfer through the sample induced by the alternating
vibration stress;\cite{NB} the corresponding relaxation rate $\tau ^{-1}$ in
Eq. (1) is directly proportional to the thermal conductivity, which is
generally a slowly varying function of temperature, so that a very broad
peak results. The observed lowering of dissipation after the high
temperature treatments should be mainly explained in terms of a change of
the charge carrier concentration and mobility, which in turn affects the
thermal conductivity. During the present experiments, however, no relevant
changes of the electrical conductivity were observed. Therefore, it is
unlikely that the main contribution to the background dissipation in the
as-grown samples was of thermoelastic origin.

Dislocations are formed during the crystal growth and subsequent cooling,
and their stress-induced movement may produce a broad background
dissipation. The decrease of such a dissipation after the high temperature
treatments can be caused by the migration of defects onto the dislocation
lines, resulting in the pinning of the dislocations.\cite{Sch63} The
migrating defects should be rather mobile, since the lowering of dissipation
starts already a few tens of kelvin above room temperature; they can be
oxygen or hydrogen atoms. It should be noted that a few at ppm impurities
are sufficient to saturate the whole dislocation net.

The presence of the peak at 400$~$K in the semiconducting sample 3 and its
disappearance during its measurement, which occurs at a temperature at which
there is certainly no H loss, indicates that the sample is not in thermal
equilibrium. The species giving rise to the process is again rather mobile;
therefore, the peak is likely due to a H-related defect, but it is not
possible at present to make any attribution.

\subsection{Interpretation of peak A and proposition of a model for the
semi-insulating state}

From the above results it appears that peak A is present only in the SI
state of both Fe-free and Fe-doped InP, suggesting that there is a common
mechanism of conversion to the SI state, which is also responsible for the
anelastic relaxation process.

The SI conversion in InP:Fe is attributed to the removal of the shallow
donors by the deep acceptor Fe atoms. Similarly, during the high temperature
treatments in P atmospheres necessary to reach the SI state in undoped InP,
the activation of some sort of deep acceptors able to compensate the shallow
donors introduced by impurities has been unanimously invoked. However, there
is no agreement at present on the nature of such acceptors and consequently
on the SI conversion mechanism. Bliss {\it et al.}\cite{BBG95} reported that
the 2316~cm$^{-1}$ line attributed to V$_{\text{In}}-$H$_{4}$ disappears
after the TT finalized to the SI conversion, and in addition they observed
that Fe and Cu were introduced during their TT. On this basis they
concluded, in accordance with Hirt {\it et al.},\cite{HWM93} that the high
resistivity values are due to the contamination by Fe and Cu deep acceptors
during the annealing process and to the concomitant outdiffusion of the V$_{%
\text{In}}-$H$_{4}$ complexes acting as intrinsic donors. We exclude that Fe
plays such a role in our undoped SI samples, since according to the chemical
analysis sample 1 contained less than 1 atomic part per billion, and the TT
for the SI conversion was carried out under controlled purity conditions.

For the model we presently propose we assume that hydrogen and vacancies are
abundant, even though not intentionally introduced. As discussed in the
following paragraph, we suppose that the as-grown material mainly contains V$%
_{\text{In}}-$H$_{4}$ complexes, whereas lower order complexes are also
present if H is trapped by other defects, like Fe. This assumption is
corroborated by several observations\cite{DPR93,TCC88} of the LVM at 2316~cm$%
^{-1}$ (V$_{\text{In}}-$H$_{4}$) in as-grown InP, and of the 2202~cm$^{-1}$
(V$_{\text{In}}-$H) line in Fe-doped\cite{TCC88} and proton-implanted\cite
{FMM92} InP. We believe that the disappearance of the V$_{\text{In}}-$H$_{4}$
line\cite{RNS88} consequent to the TT at 950$~^{\text{o}}$C for the SI
conversion is due to the outdiffusion of H, rather than of the whole V$_{%
\text{In}}-$H$_{4}$ complex, since we expect that V$_{\text{In}}$ is less
mobile than H. As a consequence of the H loss from the sample,
transformation of V$_{\text{In}}-$H$_{4}$ complexes into complexes with $n<4$
takes place. Indeed, after thermal annealings above 600$~^{\text{o}}$C new
lines appear at lower-energy,\cite{DPR93} attributed to V$_{\text{In}}-$H$%
_{n}$ with $n<4.$

Considering that V$_{\text{In}}$ is a triple acceptor,\cite
{BBG95,SVP94,EOJ96} we expect that the V$_{\text{In}}-$H$_{3}$ complex is
neutral, as the H atoms supply the three electrons. Similarly, V$_{\text{In}%
}-$H$_{2}$ and V$_{\text{In}}-$H are expected to be acceptor complexes, and
we hypothesize that they are deep acceptors. The dissociation of the V$_{%
\text{In}}-$H$_{4}$ complexes during high temperature annealing reduces the
number of shallow donors, and the newly formed V$_{\text{In}}-$H$_{n}$ $%
(n=0-2)$ deep acceptors remove the free carriers introduced by impurities
and render the material semi-insulating.

The proposed model is congruent with the observation that peak A (attributed
to V$_{\text{In}}-$H$_{n}$ with $n=1-3$) in undoped InP appears only after
the TT at 950$~^{\text{o}}$C, which transforms the complexes with $n=4$ into
lower order complexes; therefore, the existence of the peak should be
closely related to the SI state. Accordingly, we expect that peak A is
absent in as-grown undoped samples, as experimentally observed (Fig.4). The
occurrence of the peak also in the Fe-doped sample, which did not undergo
the TT, can be explained in terms of partial trapping of H in Fe-H pairs,
with consequent transformation of V$_{\text{In}}-$H$_{4}$ centres into lower
order complexes.

\subsection{Formation of V-H$_{n}$ complexes}

It remains to be explained why InP prepared by the LEC method contains a
substantial concentration of V-H$_{n}$ complexes, as indicated by the
corresponding LVM. In fact, the equilibrium concentration of vacancies in
crystals are very small even near the melting temperature, and for the case
of InP it seems natural that vacancies of the more volatile P are
predominant over the V$_{\text{In}}$. However, the formation energy of V$_{%
\text{In}}$ can be lowered if the P dangling bonds are saturated by H atoms,
and could even become negative; then, in the presence of H the formation of
In vacancies saturated by H would be possible or even energetically
favorable. During the LEC crystal growth, hydrogen would be supplied by the
capping liquid, which contains H$_{2}$O;\cite{Cle91} H could dissolve in the
liquid InP forming P-H bonds and then be incorporated in the crystal as V$_{%
\text{In}}$-H$_{4}$.

That the P-H bonds are much more stable than the In-H bonds is demonstrated
by the lack of In-H LVM in InP even after the irradiation with H, which
produces both P and In vacancies and self-interstitials. Within the proposed
picture, the high stability of the P-H bond is consistent with the stability
of peak A against vacuum annealings up to the sample decomposition (Fig. 4)
and with the relatively high barrier for the hopping of H between different
P bonds in the V$_{\text{In}}$\ (0.67$~$eV, compared to 0.22$~$eV for the
reorientation of the B-H complex in Si).

The formation of superabundant vacancies is already observed in metal
hydrides at high temperatures,\cite{Fuk95a} when the binding energy of H to
the metal vacancies can substantially lower the V-H$_{n}$ formation energy.
By exposing the metal to a high H pressure at high temperature, it is
possible to let very high concentrations of H-saturated vacancies to migrate
into the bulk (even tens of percent).

\section{Conclusion}

Anelastic spectroscopy experiments on InP crystals grown by the liquid
encapsulated Czochralski method have revealed the presence of defects whose
reorientation rate is thermally activated over a barrier of 0.67$~$eV. The
corresponding anelastic relaxation process is observed only in the
semi-insulating InP samples, irrespective of whether the semi-insulating
state is obtained by doping with Fe or by annealing near 900$~^{\text{o}}$C.
It is proposed that these defect are In vacancies with the P dangling bonds
partially saturated by H, whereas vacancies completely filled with four H
atoms do not cause anelastic relaxation. A possible mechanism for the
formation of the V$_{\text{In}}$-H$_{4}$ complexes and for the reversion to
the semi-insulating state after annealing near 900$~^{\text{o}}$C is
proposed, based on the stability of the P-H bonds. The formation of such
bonds with H dissolved from the capping liquid during the crystal growth
would make the formation of V$_{\text{In}}$-H$_{4}$ complexes favorable..
Such complexes would be donors which contribute to make the crystal a
semiconductor. The subsequent treatment at 900$~^{\text{o}}$C would
partially outgas H from the sample and remove it from the V$_{\text{In}}$-H$%
_{4}$ complexes, transforming them into neutral or deep acceptor defects.


\section{Captions to figures}

Fig. 1 Elastic energy loss versus temperature of semi-insulating InP. Sample
1: highly pure and subjected to the thermal treatment for conversion to the
SI state; sample 2: doped with 10$^{16}$~cm$^{-3}$ Fe.

Fig. 2 Peak A of sample 2 during a series vacuum annealings for about 30 min
at temperatures subsequently increased from 400 and 660$~$K at steps of 40$~$%
K.

Fig. 3 Peak A in sample 2 after the 7$^{\text{th}}$ vacuum annealing at 620$%
~ $K, after subtraction of the background. The continuous lines are single
time Debye curves with $E=0.67~$eV and $\tau _{0}^{-1}=3\times 10^{14}$~s$%
^{-1}$.

Fig. 4 Elastic energy absorption of semiconducting sample 3 ($f=2$~kHz)
during two subsequent measurements (curves 1 and 2). For comparison, the
dissipation curves of the SI samples 1 (2~kHz)\ and sample 2 (1.2~kHz) are
also shown.

Fig. 5 Evolution of the anelastic spectrum of semiconducting InP (sample 3)
measured in vacuum: as prepared (curve 1); 2$^{\text{nd}}$ heating run
(curve 2); after subsequent annealings in vacuum for 30 min at the
temperatures indicated in the figure.

Fig. 6 Elastic energy loss of SI InP:Fe (sample 2) and of non-SI InP (sample
3) after the last vacuum annealings at 660 and 650$~$K respectively

Fig. 7 Peak A in sample 2 after annealings of 30~min in vacuum at the
increasing temperatures indicated in the legend; these measurements follow
those of Fig. 2.

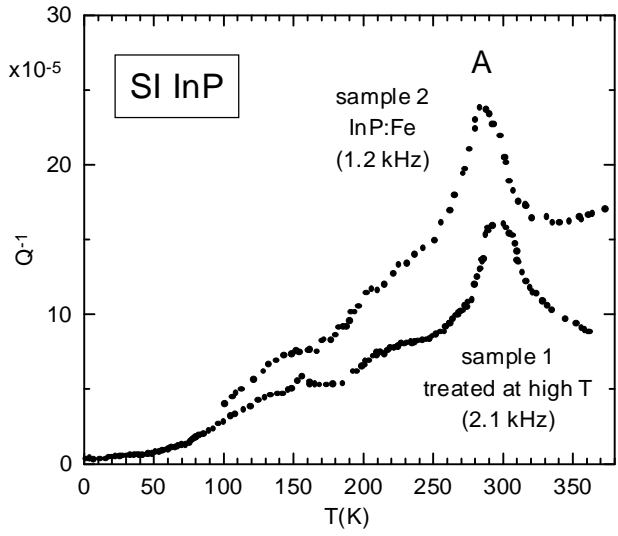
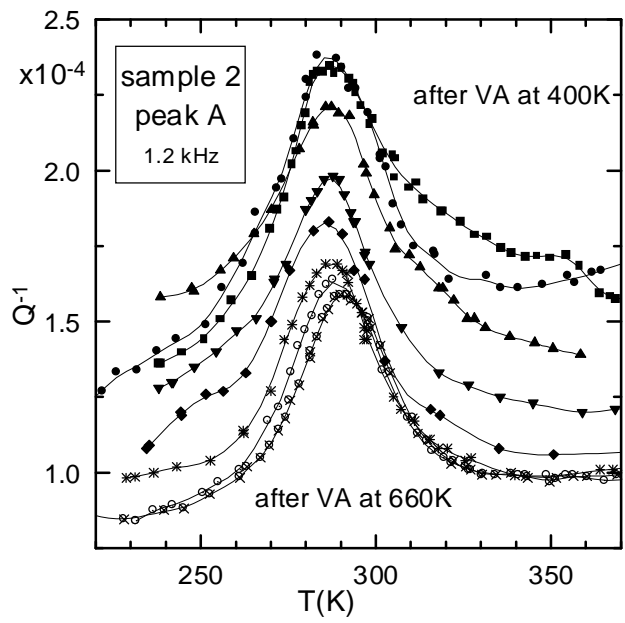
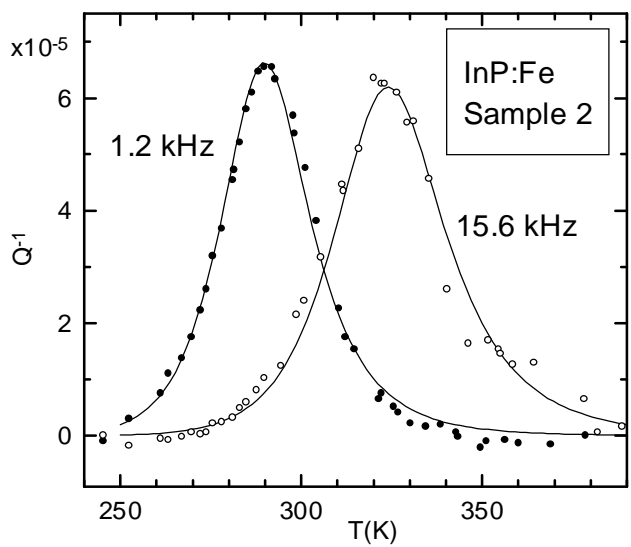
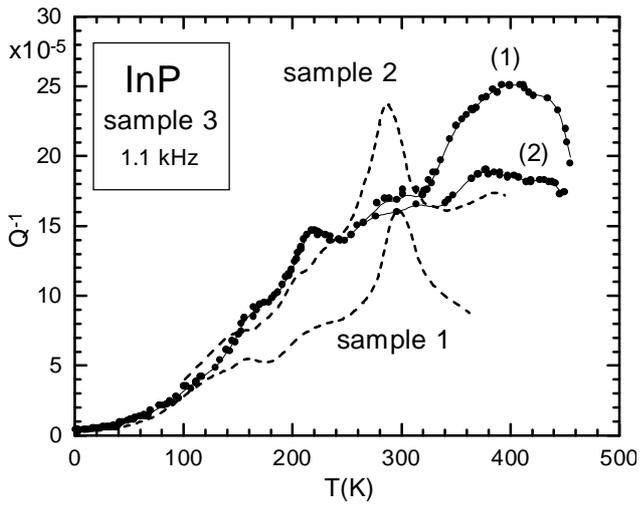
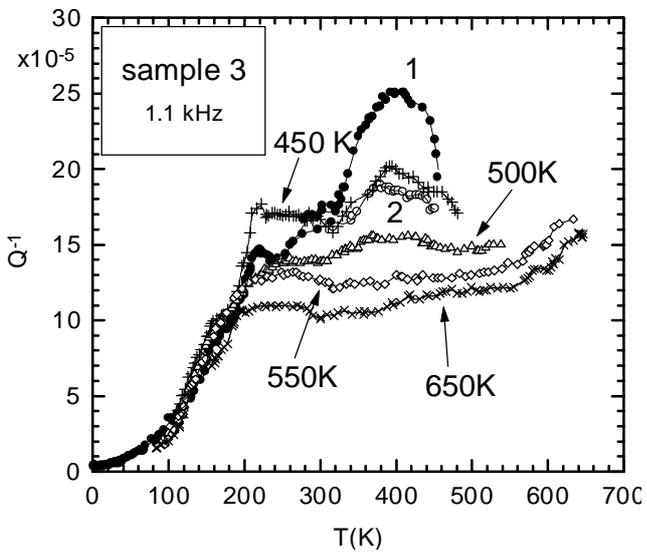
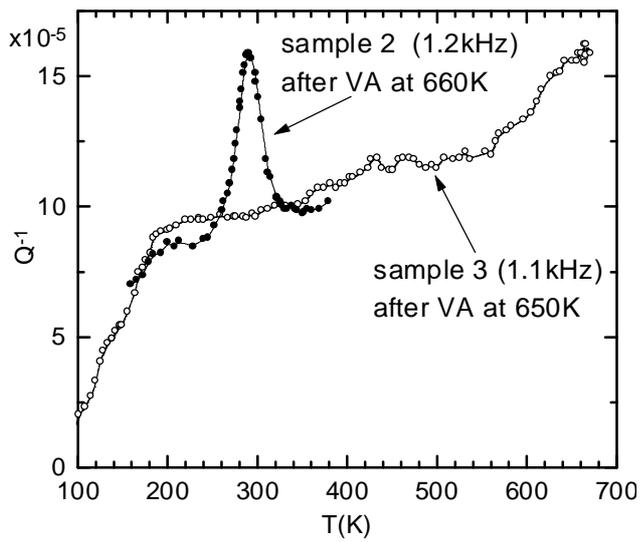
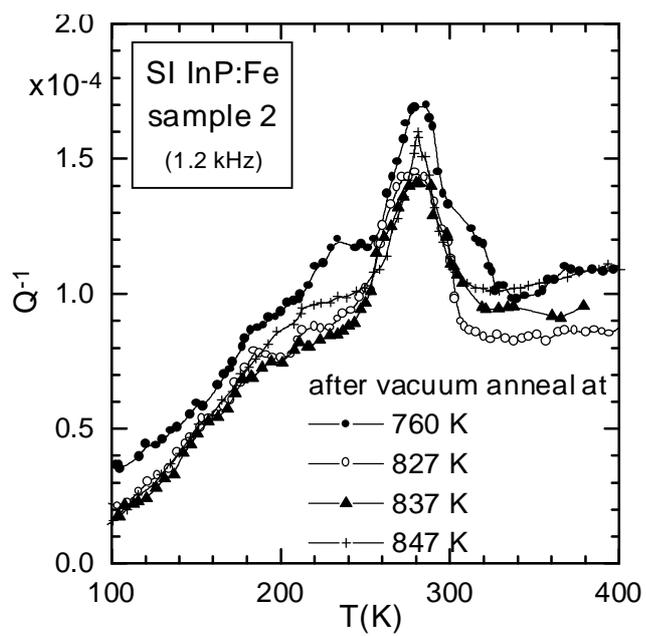
\end{document}